\documentclass[10pt, conference]{IEEEtran}

\usepackage{theorem}
\usepackage[utf8]{inputenc} 
\usepackage[T1]{fontenc}
\usepackage{url}
\usepackage{ifthen}
\usepackage{cite}
\usepackage[cmex10]{amsmath} 
\usepackage{amssymb}
\usepackage{mathtools}
\usepackage{mathrsfs}
\usepackage{multirow}
\usepackage{authblk}
\usepackage{tikz}
\usepackage{adjustbox}
\usetikzlibrary{graphs}
\usetikzlibrary{positioning}

\newtheorem{theorem}{Theorem}
\newtheorem{lemma}{Lemma}

\newtheorem{remark}{Remark}
\newtheorem{definition}{Definition}

\newenvironment{Proof}[1]{\medskip\par\noindent{\bf Proof:\,}\,#1}{{\mbox{\,$\blacksquare$}\par}}
\newenvironment{Proof of Lemma 1}[1]{\medskip\par\noindent{\bf Proof:\,}\,#1}{{\mbox{\,$\blacksquare$}\par}}
\newenvironment{Proof of Theorem 2}[1]{\medskip\par\noindent{\bf Proof:\,}\,#1}{{\mbox{\,$\blacksquare$}\par}}

\newcommand\nc\newcommand
\nc{\cA}{\mathcal{A}}\nc{\cB}{\mathcal{B}}\nc{\cC}{\mathcal{C}}\nc{\cD}{\mathcal{D}}
\nc{\cE}{\mathcal{E}}\nc{\cF}{\mathcal{F}}\nc{\cG}{\mathcal{G}}\nc{\cH}{\mathcal{H}}
\nc{\cI}{\mathcal{I}}\nc{\cJ}{\mathcal{J}}\nc{\cK}{\mathcal{K}}\nc{\cL}{\mathcal{L}}
\nc{\cM}{\mathcal{M}}\nc{\cN}{\mathcal{N}}\nc{\cO}{\mathcal{O}}\nc{\cP}{\mathcal{P}}
\nc{\cQ}{\mathcal{Q}}\nc{\cR}{\mathcal{R}}\nc{\cS}{\mathcal{S}}\nc{\cT}{\mathcal{T}}
\nc{\cU}{\mathcal{U}}\nc{\cV}{\mathcal{V}}\nc{\cW}{\mathcal{W}}\nc{\cX}{\mathcal{X}}
\nc{\cY}{\mathcal{Y}}\nc{\cZ}{\mathcal{Z}}

\nc{\bba}{\mathbf{a}}\nc{\bbb}{\mathbf{b}}\nc{\bbc}{\mathbf{c}}\nc{\bbd}{\mathbf{d}}
\nc{\bbe}{\mathbf{e}}\nc{\bbf}{\mathbf{f}}\nc{\bbg}{\mathbf{g}}\nc{\bbh}{\mathbf{h}}
\nc{\bbi}{\mathbf{i}}\nc{\bbj}{\mathbf{j}}\nc{\bbk}{\mathbf{k}}\nc{\bbl}{\mathbf{l}}
\nc{\bbm}{\mathbf{m}}\nc{\bbn}{\mathbf{n}}\nc{\bbo}{\mathbf{o}}\nc{\bbp}{\mathbf{p}}
\nc{\bbq}{\mathbf{q}}\nc{\bbr}{\mathbf{r}}\nc{\bbs}{\mathbf{s}}\nc{\bbt}{\mathbf{t}}
\nc{\bbu}{\mathbf{u}}\nc{\bbv}{\mathbf{v}}\nc{\bbw}{\mathbf{w}}\nc{\bbx}{\mathbf{x}}
\nc{\bby}{\mathbf{y}}\nc{\bbz}{\mathbf{z}}

\nc{\bbA}{\mathbf{A}}\nc{\bbB}{\mathbf{B}}\nc{\bbC}{\mathbf{C}}\nc{\bbD}{\mathbf{D}}
\nc{\bbE}{\mathbf{E}}\nc{\bbF}{\mathbf{F}}\nc{\bbG}{\mathbf{G}}\nc{\bbH}{\mathbf{H}}
\nc{\bbI}{\mathbf{I}}\nc{\bbJ}{\mathbf{J}}\nc{\bbK}{\mathbf{K}}\nc{\bbL}{\mathbf{L}}
\nc{\bbM}{\mathbf{M}}\nc{\bbN}{\mathbf{N}}\nc{\bbO}{\mathbf{O}}\nc{\bbP}{\mathbf{P}}
\nc{\bbQ}{\mathbf{Q}}\nc{\bbR}{\mathbf{R}}\nc{\bbS}{\mathbf{S}}\nc{\bbT}{\mathbf{T}}
\nc{\bbU}{\mathbf{U}}\nc{\bbV}{\mathbf{V}}\nc{\bbW}{\mathbf{W}}\nc{\bfX}{\mathbf{X}}
\nc{\bbY}{\mathbf{Y}}\nc{\bbZ}{\mathbf{Z}}

\nc{\sA}{\mathsf{A}}\nc{\sB}{\mathsf{B}}\nc{\sC}{\mathsf{C}}\nc{\sD}{\mathsf{D}}
\nc{\sE}{\mathsf{E}}\nc{\sF}{\mathsf{F}}\nc{\sG}{\mathsf{G}}\nc{\sH}{\mathsf{H}}
\nc{\sI}{\mathsf{I}}\nc{\sJ}{\mathsf{J}}\nc{\sK}{\mathsf{K}}\nc{\sL}{\mathsf{L}}
\nc{\sM}{\mathsf{M}}\nc{\sN}{\mathsf{N}}\nc{\sO}{\mathsf{O}}\nc{\sP}{\mathsf{P}}
\nc{\sQ}{\mathsf{Q}}\nc{\sR}{\mathsf{R}}\nc{\sS}{\mathsf{S}}\nc{\sT}{\mathsf{T}}
\nc{\sU}{\mathsf{U}}\nc{\sV}{\mathsf{V}}\nc{\sW}{\mathsf{W}}\nc{\sX}{\mathsf{X}}
\nc{\sY}{\mathsf{Y}}\nc{\sZ}{\mathsf{Z}}

\IEEEoverridecommandlockouts
\allowdisplaybreaks

\begin{document}
\title{Private Information Retrieval on Multigraph-Based Replicated Storage} 

\author[1]{Shreya Meel}
\author[2]{Xiangliang Kong}
\author[1]{Thomas Jacob Maranzatto}
\author[2]{Itzhak Tamo}
\author[1]{Sennur Ulukus}

\affil[1]{\normalsize University of Maryland, College Park, MD, USA}
\affil[2]{\normalsize Tel Aviv University, Tel Aviv-Yafo}
\affil[1]{\normalsize \{\textit{smeel, tmaran, ulukus}\}\textit{@umd.edu}}
\affil[2]{\normalsize \textit{rongxlkong@gmail.com, tamo@tauex.tau.ac.il} }

    

\maketitle

\begin{abstract}
   We consider the private information retrieval (PIR) problem for a multigraph-based replication system, where each set of $r$ files is stored on two of the servers according to an underlying $r$-multigraph. Our goal is to establish upper and lower bounds on the PIR capacity of the $r$-multigraph. Specifically, we first propose a construction for multigraph-based PIR systems that leverages the symmetry of the underlying graph-based PIR scheme, deriving a capacity lower bound for such multigraphs. Then, we establish a general upper bound using linear programming, expressed as a function of the underlying graph parameters. Our bounds are demonstrated to be tight for PIR systems on multipaths for even number of vertices.
\end{abstract}

\section{Introduction}
Introduced in \cite{chor}, private information retrieval (PIR) is the problem of retrieving a specific file in a database from a server, without revealing the identity of the file to the server. While PIR was extensively studied in the computer science community, it is first approached from an information-theoretic perspective by Sun and Jafar in \cite{SJ17} within a non-colluding, multi-server setting. This framework adopts the traditional Shannon theoretic formulation, where the message lengths are allowed to be arbitrarily larger than the number of messages, due to which the download cost dominates the upload cost. Consequently, the \emph{rate} of PIR is formulated as the number of desired message bits retrieved per bit of download. The supremum of all achievable rates, referred to as the \emph{capacity}, is derived in \cite{SJ17} under the fully replicated setting.

Besides the original setting, PIR has been studied under more relaxed assumptions. For example, PIR without the non-colluding assumption, where servers can share their queries to learn the file index, is studied in \cite{colluding, coded_colluding_2017, mdstpir}. PIR under various threat models, such as the presence of eavesdroppers and Byzantine servers, is investigated in \cite{banawan_eaves, nan_eaves, byzantine_tpir}. The impact of side information available to the user, such as knowledge of database contents \cite{tpir_sideinfo, kadhe_singleserver_pir}, and the benefits of cache-aided PIR in achieving improved rates \cite{wei_banawan_cache_pir}, are also explored. Another important variant of PIR that received much attention is the symmetric PIR, \cite{c_spir} which additionally demands database privacy against the user. This resulted in a stream of works, such as \cite{wang_spir, tspir_mdscoded}. We refer the readers to \cite{ulukusPIRLC} for a more comprehensive survey on PIR and its variants.

All the aforementioned PIR systems consider fully replicated storage, which may be difficult to achieve in practice. This motivated the study of PIR with non-replicated servers in \cite{graphbased_pir} through a graph-based model. In this model, a graph/hypergraph represents a PIR system, where each vertex corresponds to a server, and each edge/hyperedge represents a file replicated across the servers associated with that edge/hyperedge. The authors of \cite{graphbased_pir} focus on PIR over simple graphs and study the robustness of the corresponding PIR system under server collusion. Graph-based PIR is further explored by \cite {BU19}, where bounds on the capacity of some regular graphs are proposed and the exact PIR capacity for cycle graphs is determined. In \cite{asymp_gxstpir}, the authors study the capacity of graph/hypergraph-based PIR when the number of files goes to infinity, with additional server collusion and secure storage constraints. The authors of \cite{SGT23} develop novel upper bounds on PIR capacity and propose improved PIR scheme constructions several classes of graphs. However, the PIR capacity for general graphs remains unknown except for some small graphs. For instance, the PIR capacity of the star graph is unknown beyond the case of $4$ files, with the case of 4-files being solved recently in \cite{YJ23}.

Note that the structure of a graph determines the number of files stored in each server. Specifically, it restricts the number of servers containing the same file to exactly $2$ and sets the number of files stored in a server to the degree of the corresponding vertex in the underlying graph. Although \cite{asymp_gxstpir} provides insights into scenarios where an arbitrary number of files are stored in each server, their bounds are asymptotic and do not explicitly account for the actual graph structure.

In this work, we aim to go beyond the above limitations of PIR systems modeled by graphs by investigating PIR systems over multigraphs with finite uniform multiplicity $r\geq 2$. In an $ r $-multigraph, each edge of the underlying simple graph is replaced by $ r $ multiedges. Consequently, the servers sharing a single file in a graph-based PIR system share a set of $ r $ files in a multigraph-based PIR system. 

As in prior works, our focus lies on deriving bounds for the capacity of multigraph-based PIR systems and constructing schemes that approach these bounds. First, we propose a construction based on any graph-based PIR scheme that satisfies certain properties, and derive a lower bound on the capacity of multigraphs for which such schemes exist. Next, we establish a general upper bound using linear programming. Our bounds are shown to be tight in the case of multi-paths with an even number of vertices.

\section{Problem Formulation}
Let $\cS=\{S_1,S_2,\ldots,S_N\}$ denote $N$ non-colluding servers and $\cW=\{W_1,W_2,\ldots,W_K\}$ denote $K$ independent files. Each $W_{i}\in \mathbb{F}_2^{L}$ is a binary vector of length $L$ chosen uniformly at random from $\mathbb{F}_2^{L}$, thus $H(W_i)=L$ for every $i\in [K]$, and
\begin{align}
    H(W_1,\ldots,W_K)&=H(W_1)+\cdots+H(W_K)=KL,
    \label{eq1_problem_setting}
\end{align}
{where $H(\cdot)$ is the binary entropy function.} In the PIR problem, a user privately generates $\theta\in [K]$ and wishes to retrieve $W_{\theta}$ while keeping $\theta$ a secret from each server. Let $\cQ\triangleq\{Q_i=Q_i^{(\theta)}:i\in [N]\}$ denote the set of all queries generated by the user. Since the user has no information on the content of the files, the queries are independent of them, which means that
\begin{align}\label{eq2_problem_setting}
    I(W_1,\ldots,W_K;Q_1,\ldots,Q_N)=0.
\end{align}

Upon receiving its query $Q_i$ from the user, server $S_i$ responds with an answer $A_i=A_i^{(\theta)}$, which is a function of $Q_i$ and the files it holds $\cW_{S_i}$. Thus, for any $i\in [N]$, 
\begin{align}\label{eq3_problem_setting}
    H(A_i|Q_{i},\cW_{S_i})=0.
\end{align}

A PIR scheme has two formal requirements, \emph{reliability} and \emph{privacy}. For reliability, the user should be able to retrieve the desired file $W_{\theta}$ from the received answers $A_i, i\in[N]$ with zero probability of error, hence
\begin{align}
&\textbf{[Reliability]} \quad H(W_\theta|A_{[N]},Q_{[N]})=0.\label{eq4_problem_setting}\\
    \intertext{\indent For privacy, each server learns no information about the desired file index $\theta$. That is, for any $i\in [N]$ and $\theta\in [K]$,}&\textbf{[Privacy]} \quad (Q_i^{(1)},A_{i}^{(1)},\cW_{S_i})\sim (Q_i^{(\theta)},A_{i}^{(\theta)},\cW_{S_i}).\label{eq5_problem_setting}
\end{align}

Assume that, the $K$ files are stored such that every pair of servers either share a set of $r$ distinct files, or no files at all. By relabeling the files, we can write $\cW$ as
\begin{align}
\{W_{1,1}, \ldots, W_{1,r},\ldots, W_{K',1},\ldots,W_{K',r}\},
\end{align}
and index each file by a pair $(i,j)\in [K']\times [r]$, where $K'=K/r$. Then, this model can naturally be described by an $r$-multigraph $ G^{(r)} = (\cS, \cW) $, whose vertex set is the set of servers $ \cS $, and an $r$-subset of files $ \{W_{i,1},\ldots,W_{i,r}\} \in \cW $ is viewed as the $r$-multiedge between $S_i$ and $S_j$ if $W_{i,1},\ldots,W_{i,r}$ are stored on servers $ S_i $ and $ S_j $. We denote by $ \cW_{S_i} $ the set of files stored on server $ S_i $, and by $ \cW_{i,j} $ the set of $r$ files stored on servers $ S_i $ and $ S_j $. Hence, an instance of the PIR problem is uniquely defined by a $r$-multigraph, and any $r$-multigraph gives rise to an instance of the above PIR problem. Specially, when $r=1$, it reduces to the PIR problem over simple graphs studied in \cite{graphbased_pir,BU19,asymp_gxstpir,SGT23}.

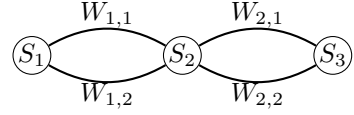
\begin{figure}[h]
    \centering
    \begin{tikzpicture}[every node/.style={circle, draw, inner sep=0.3pt, minimum size=0.15cm}, 
    every edge/.style={draw, thick}, 
    edge label/.style={draw=none, rectangle, inner sep=0pt}]
    
    \node (A) at (0, 0) {$S_1$};
    \node (B) at (2, 0) {$S_2$};
    \node (C) at (4, 0) {$S_3$};
    
    \draw[thick, bend right=30] (A) to node[edge label, below] {$W_{1,2}$} (B);
    \draw[thick, bend left=30] (A) to node[edge label, above] {$W_{1,1}$} (B);
    
    \draw[thick, bend right=30] (B) to node[edge label, below] {$W_{2,2}$} (C);
    \draw[thick, bend left=30] (B) to node[edge label, above] {$W_{2,1}$} (C);
    
    \end{tikzpicture}
    \vspace{-0.15cm}
    \caption{The PIR system corresponding to $\mathbb{P}_3^{(2)}$.}
    \label{fig1}
\end{figure}

For example, Fig.~\ref{fig1} illustrates a model based on a $2$-multi-path over $3$ vertices, which consists of three servers $ \cS = \{S_1, S_2, S_3\} $ and four files $ \cW = \{W_{1,1}, W_{1,2}, W_{2,1}, W_{2,2}\} $. Accordingly, server $S_1$ and $S_3$ store exactly two files, i.e., $ \cW_{S_1} = \{W_{1,1}, W_{1,2}\} $, $ \cW_{S_3} = \{W_{2,1}, W_{2,2}\} $, and server $S_2$ stores all the four files, i.e., $ \cW_{S_2} = \cW $.

Throughout the paper, we use $ G = (V, E) $ to denote a simple graph with vertex set $ V $ and edge set $ E $. We use $ G^{(r)} $ to denote the multigraph extension of the graph $ G $ with multiplicity $ r $. The PIR capacity of the multigraph $G^{(r)}$ is defined as,
\begin{align}
    \mathscr{C}\left(G^{(r)}\right) \triangleq \sup_{T'} \frac{L}{\sum_{i=1}^N H(A_i)},
\end{align}
where the supremum is taken over all possible schemes $T'$ on $G^{(r)}$, assuming the file size $L$ can be arbitrarily large.

\section{A Lower Bound on the PIR Capacity}\label{Sec: Scheme Constructions}

\subsection{Motivating Example}\label{example:r_multipath}
\begin{figure}[h]
    \centering
    \begin{tikzpicture}[
        every node/.style={circle, draw, inner sep=0.3pt, minimum size=0.15cm}, 
        every edge/.style={draw, thick}, 
        edge label/.style={draw=none, rectangle, inner sep=2pt, font=\small}
    ]
    \node (A) at (0, 0) {$S_1$};
    \node (B) at (1.75, 0) {$S_2$};
    \node (C) at (3.5, 0) {$S_3$};

    \draw[thick] (A) to node[edge label, below] {$W_{1}$} (B);

    \draw[thick] (B) to node[edge label, below] {$W_{2}$} (C);
    \end{tikzpicture}
    \caption{The PIR system corresponding to $\mathbb{P}_3$}
    \label{fig2}
\end{figure}
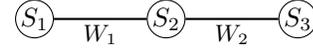

We motivate the underlying idea of our general multigraph scheme using the example of the PIR scheme for the path graph $ \mathbb{P}_3 $, which consists of $3$ servers, $ S_1 $, $ S_2 $, $ S_3 $ and $2$ files. As illustrated in Fig.~\ref{fig2}, the files are stored as follows: $ W_1 $ at $ S_1 $, $\{ W_1 ,W_2\} $ at $ S_2 $, and $ W_2 $ at $ S_3 $.

Suppose that each file consists of two bits, and the user wants to privately retrieve $ W_{\theta} $. To do so, the user privately and independently permutes the bits of each $W_i$, $i\in [2]$, and the corresponding permuted file is denoted as ${w}_i$. Let us write $ {w}_1 = (a_1, a_2) $ and $ {w}_2 = (b_1, b_2) $. The user sends queries to retrieve three answer bits from  $ S_1 $, $ S_2 $ and $ S_3 $. The answers downloaded from the servers for $ \theta=1, 2 $ are given in Table~\ref{tab:base_answers}. Clearly, the user can decode the desired file in both cases. Also, since the permutation of bits is unknown to the servers, the queries appear uniformly distributed, regardless of the value of $ \theta $.

\begin{table}[h]
    \centering
    \caption{Answer table for $\mathbb{P}_3$.}
    \label{tab:base_answers}
    \begin{tabular}{|c|c|c|c|}
        \hline
         & $S_1$ & $S_2$ & $S_3$\\
        \hline
        $\theta=1$ & $a_1$ & $a_2+b_2$ & $b_2$\\
        \hline
        $\theta = 2$ & $a_1$ & $a_1+b_1$ & $b_2$\\
        \hline
    \end{tabular}
\end{table}

Now, consider the PIR system corresponding to the multigraph $ \mathbb{P}_3^{(2)} $, which is illustrated in Fig.~\ref{fig1}. Suppose each file consists of 4 bits. Let the message bits, after the user's private permutation, be denoted as,
\begin{align}
    & w_{1,1} = (a_1, a_2, a_3, a_4), & & {w}_{1,2} = (a'_1, a'_2, a'_3, a'_4), \nonumber \\
    & {w}_{2,1} = (b_1, b_2, b_3, b_4), & & {w}_{2,2} = (b'_1, b'_2, b'_3, b'_4).
\end{align}
Then, the answers from each server for different queries are shown in Table~\ref{tab:multipath_r=2}. For each query $ \theta = (i,j) $, each server responds with three answer bits. The first answer bit is identical to that in the above scheme for $ \mathbb{P}_3 $ with $ \theta = i $ and files $ \{W_{1,1}, W_{2,1}\} $, and the second answer bit is identical to that in the above scheme for $ \mathbb{P}_3 $ with $ \theta = i $ and files $ \{W_{1,2}, W_{2,2}\} $. The third answer bit is constructed by applying the scheme of $ \mathbb{P}_3 $ to carefully-designed sums of message bits.

\begin{table}[h]
    \centering
    \vspace{-0.2cm}
    \caption{Answer table for $\mathbb{P}_3^{(2)}$.}
    \label{tab:multipath_r=2}
 \begin{tabular}{|c|c|c|c|}   
 \hline
 & $S_1$ & $S_2$ & $S_3$\\
 \hline
 \multirow{3}{*}{$\theta = (1,1)$} & $a_1$ & $a_2+b_2$ & $b_2$\\
 & $a_1'$ & $a_2'+b'_2$ & $b'_2$\\
 & $a_4+a_2'$ & $a_3+a'_1+b_4+b'_4$ & $b_4+b'_4$\\
 \hline
  \multirow{3}{*}{$\theta = (1,2)$} & $a_1$ & $a_2+b_2$ & $b_2$\\
 & $a_1'$ & $a_2'+b'_2$ & $b'_2$\\
 & $a_2+a_4'$ & $a_1+a'_3+b_4+b'_4$ & $b_4+b'_4$\\
 \hline
  \multirow{3}{*}{$\theta = (2,1)$} & $a_1$ & $a_1+b_1$ & $b_2$\\
 & $a_1'$ & $a_1'+b'_1$ & $b'_2$\\
 & $a_2+a_2'$ & $a_2+a'_2+b_4+b'_2$ & $b_3+b'_1$\\
 \hline
  \multirow{3}{*}{$\theta = (2,2)$} & $a_1$ & $a_1+b_1$ & $b_2$\\
 & $a_1'$ & $a_1'+b'_1$ & $b'_2$\\
 & $a_2+a_2'$ & $a_2+a'_2+b_2+b'_4$ & $b_1+b'_3$\\
 \hline
 \end{tabular}
\end{table}

For example, consider the case when $ \theta = (1,1) $. By the first and second answer bits from all $3$ servers, the user can retrieve $a_1, a_2$ and $a_1',a_2'$, respectively. The third answer bit from $ S_1 $ is a new message bit of $ W_{\theta} $, i.e., $ a_4 $ added to $ a_2' $, where the latter is available to the user. From $ S_3 $, the third answer bit is the sum of unused message bits of $ W_{2,1} $ and $ W_{2,2} $, i.e., $ b_4+b_4' $. The third answer bit from $ S_2 $ is the sum of bits from all $4$ files stored in $S_2$, comprising a new message bit  $ a_3 $ of $ W_{\theta} $, $ a_1' $ and $ b_4+ b_4'$, where $ a_1' $ is retrieved from the second answers of all servers, and $ b_4+ b_4'$ is the third answer bit from $S_3$. This way, all four bits of $ W_{\theta} $ are retrieved.

To understand why privacy holds, note that the query structure seen by each server is identical for all $ \theta $. Moreover, the private permutations of message bits, generated by the user, hide the actual bit indices from each server.

\subsection{General Scheme for $r$-Multigraph}
Before stating Theorem \ref{thm:achievable_lbnd}, we need the following definition.

\begin{definition}[Symmetric Retrieval Property]
A graph-bas-ed PIR scheme is said to satisfy the symmetric retrieval property (SRP) if, for every possible $ \theta $, the number of bits of the file $W_{\theta}$ retrieved from each of the servers storing $ W_{\theta} $, is equal. Equivalently,
    $H(A_{i}^{(\theta)}|Q_{i}^{(\theta)},\cW_{i}\setminus\{W_{\theta}\})=H(A_{j}^{(\theta)}|Q_{j}^{(\theta)},\cW_{j}\setminus\{W_{\theta}\})=\frac{H(W_{\theta})}{2},
$ if $W_{\theta}$ is replicated on servers $S_i$ and $S_j$, which store the set of files $\mathcal{W}_{S_i}$ and $\mathcal{W}_{S_j}$, respectively.
\end{definition}

\begin{theorem}\label{thm:achievable_lbnd}
Let $T$ denote a scheme for the simple graph $G$, that satisfies SRP and yields the rate $R(G)$. Then, there exists a scheme $T^{(r)}$ for the corresponding $r$-multigraph $G^{(r)}$ which gives the capacity lower bound,
\begin{align}\label{eq:multigraph_achievable_rate}
    \mathscr{C}\left(G^{(r)}\right) &\geq R(G) \cdot \left(\frac{1}{2-\frac{1}{2^{r-1}}} \right).
\end{align}
\end{theorem}

\begin{Proof}
Let $T$ be the PIR scheme for the graph $G = (\cS, \cW_0)$, whose vertices are servers in $\cS$ and edges are associated with files in 
\begin{align}\label{define_files_G}
    \cW_0 = \{W_{1,0}, W_{2,0}, \ldots, W_{K',0}\}.
\end{align}
Suppose $ T $ admits $L'$ bits per file and $D'$ downloaded bits. Then, $ R(G) = \frac{L'}{D'} $. Next, we construct a PIR scheme $ T^{(r)} $ for $ G^{(r)} $ based on $ T $.

{Suppose} that each file $ W_{s,t} $, $ (s,t) \in [K'] \times [r] $, consists of $ L = 2^{r-1}L' $ bits in $ T^{(r)} $. Before sending out the queries, the user first permutes the $ L $ bits of each of the $ rK' $ files through a private mapping $ \sigma_{s,t}: [L] \rightarrow [L] $ independently. We denote $ \sigma_{s,t}(W_{s,t}) = {w}_{s,t} $. Let $ \theta = (i,j) $ be the desired file index, and assume that $ W_{i,j} $ is stored on servers $ S_{n_1} $ and $ S_{n_2} $.  Then, as in the motivating example, the scheme $ T^{(r)} $ proceeds in $ r $ stages. In each stage, we apply $T$ to several graphs on $\cS$, each isomorphic to $G$, where every edge represents a sum of some of the $r$ files stored on its two associated vertices.

In the following, to simplify the description of the retrieval process, assume without loss of generality that $\theta=(1,1)$.

In the first stage, for every $t\in [r]$, we apply $ T $ to graph $G_{\{t\}}=(\cS, \cW_{\{t\}})$, where 
\begin{align}
    \cW_{\{t\}}\triangleq\{W_{1,t},W_{2,t},\ldots,W_{K',t}\},
\end{align}
and retrieve the first $ L' $ bits of the file $W_{1,t}$ stored on $ S_{n_1} $ and $ S_{n_2} $. This requires $ rD' $ downloaded bits in total and results in 
\begin{align}
    {w}_{1,t}[1:L'],~\forall~t \in [r],
\end{align}
where $ {w}_{1,1}[1:L'] $ are the desired message bits, and $ {w}_{1,t}[1:L'] $, $ t \neq 1 $, are the interference bits used for retrieval in the second stage. By the SRP, we can adjust $ T $ so that, for each $ t \in [r] $, the first half $ {w}_{1,t}[1:L'/2] $ is retrieved from $ S_{n_1} $, and the latter half $ {w}_{1,t}[L'/2+1:L'] $ is retrieved from $ S_{n_2} $.

In the second stage, for every $ 2 $-subset $ \{j_1, j_2\} \subseteq [r] $, we apply $ T $ to graph $G_{\{j_1,j_2\}}=(\cS, \cW_{\{j_1,j_2\}})$, where
\begin{align}
    \cW_{\{j_1,j_2\}}\triangleq\{W_{i,j_1}+W_{i,j_2}:~i\in [K']\},
\end{align}
and retrieve $ L' $ bits of the file ${W}_{1,j_1} + {W}_{1,j_2}$ stored on $ S_{n_1} $ and $ S_{n_2} $. Specifically, we have:
\begin{itemize}
    \item If $1\in \{j_1,j_2\}$, assume without loss of generality that $j_1=1$. Then, we apply $T$ to retrieve
    \begin{align}
    {w}_{1,1}[(j_2-1)L'+1:j_2 L']+{w}_{1,j_2}[1:L'].
    \end{align}
    \item If $1\notin \{j_1,j_2\}$, then we apply $T$ to retrieve \begin{align}\label{interference_stage2}
    &{w}_{1,j_1}[(j_2-1)L'+1:j_2L'] \nonumber\\
    &+{w}_{1,j_2}[(j_1-1)L'+1:j_1L'].
    \end{align} 
\end{itemize}
Similarly, by SRP, we can assume that the first $L'/2$ retrieved bits of ${w}_{1,1}+{w}_{1,j_2}$ is from $S_{n_2}$ and the latter $L'/2$ bits is retrieved from $S_{n_1}$, and for each $\{j_1,j_2\}\subseteq [r]\setminus\{1\}$, the first $L'/2$ retrieved bits of ${w}_{1,j_1} + {w}_{1,j_2}$ is from $S_{n_1}$ and the latter $L'/2$ bits is retrieved from $S_{n_2}$. 

Consider the bits of $ {W}_{1,j_1} + {W}_{1,j_2} $ retrieved in the second stage. Since ${w}_{1,j_2}[1:L']$ are known by the first stage, we can retrieve $(r-1)L'$ bits of the form ${w}_{1,1}[(j_2-1)L'+1:j_2L']$, $2\leq j_2\leq r$, for the desired file. Moreover, for each $\{j_1,j_2\}\subseteq [r]\setminus\{1\}$, we also have $L'$ bits from the sum $ {W}_{1,j_1} + {W}_{1,j_2} $. These ${{r-1}\choose 2}L'$ bits are left as interference bits to be used for retrieval in the third stage.

Generally, for $\ell\geq 3$ and every $ (\ell-1) $-set $ B \subseteq [r] \setminus \{1\} $, we uniquely assign it an integer 
\begin{align}
    u_{B} \in \left[\sum_{s=0}^{\ell-2}{{r-1}\choose s}+1, \sum_{s=0}^{\ell-2}{{r-1}\choose s} + \binom{r-1}{\ell-1}\right], 
\end{align} 
and denote $U_{B} \triangleq [(u_{B} - 1)L' + 1: u_{B}L']$. Then, in the $\ell$-th stage, for every $ \ell $-subset $ A \subseteq [r] $, $T$ is applied to graph $G_{A}=(\cS,\cW_{A})$, where
\begin{align}
    \cW_{A}\triangleq \left\{\sum_{t\in A}W_{i,t}:~i\in [K']\right\},
\end{align}
to retrieve the following $L'$ bits of $\sum_{t\in A}{W}_{1,t}$, 
\begin{itemize}
    \item If $1\in A$, then we apply $T$ to retrieve
    \begin{align}
        {w}_{1,1}|_{U_{A\setminus\{1\}}}+\sum_{t\in A\setminus\{1\}}{w}_{1,t}|_{U_{A\setminus\{1,t\}}}.
    \end{align}
    \item If $1\notin A$, then we apply $T$ to retrieve \begin{align}\label{interference_stagel}
        \sum_{t\in A}{w}_{1,t}|_{U_{A\setminus\{t\}}}.
    \end{align}
\end{itemize}
By SRP, for $A$ containing $1$, the first $L'/2$ retrieved bits of $\sum_{t\in A}{w}_{1,t}$ is from $S_{n_2}$ and the latter $L'/2$ bits is retrieved from $S_{n_1}$, and for $A$ without $1$, the first $L'/2$ retrieved bits of $\sum_{t\in A}{w}_{1,t}$ are from $S_{n_1}$ and latter $L'/2$ bits are from $S_{n_2}$. 

Using the interference bits $\sum_{t\in B}{w}_{1,t}|_{U_{B\setminus \{t\}}}$ from the $(\ell-1)$-th stage, we can retrieve ${{r-1}\choose \ell-1}L'$ bits of form ${w}_{1,1}|_{U_{B}}$, for every $B\subseteq [r]\setminus \{1\}$ of size $\ell-1$. For each $A\subseteq [r]\setminus\{1\}$ of size $\ell$, the $L'$ bits from the sum $ \sum_{s\in A}{w}_{1,s}|_{U_{A\setminus\{s\}}} $ are left as interference bits to be used for retrieval in the next stage.

\emph{Reliability:} The reliability of $T^{(r)}$ follows directly by $T$.

\emph{Privacy:} The SRP of $T$ guarantees that, $L'/2$ interference bits of $\sum_{t\in B}{w}_{1,t}|_{U_{B\setminus \{t\}}}$ are retrieved from $S_{n_1}$ in the $(\ell-1)$-th stage and are used to retrieve  $L'/2$ message bits in $w_{1,1}+\sum_{t\in B}{w}_{1,t}|_{U_{B\setminus \{t\}}}$, retrieved from $S_{n_2}$ in the $\ell$th stage, and vice versa. This ensures that an equal number of bits from each file is downloaded from every server. That is, the files stored on every server are treated equally in the $r$-stage process. Thus, the privacy of $T^{(r)}$ follows from the privacy of $T$ in each stage and the random permutations on bits applied independently to the files of $\mathcal{W}$.

\emph{Rate:} In each stage $\ell\in [r]$, the user downloads $\binom{r}{\ell}D'$ bits. The resulting download cost $D$ is therefore,
\begin{align}
    D=\sum_{\ell=1}^r \binom{r}{\ell} D' = (2^r -1)D'.
\end{align}
This gives rise to the achievable rate, $\frac{L}{D} = \frac{2^{r-1}L'}{(2^r-1)D'}$, which is equal to \eqref{eq:multigraph_achievable_rate}.
\end{Proof}

\begin{remark}
    The scheme for $\mathbb{P}_3$ in Section \ref{example:r_multipath} can be extended to general $\mathbb{P}_N$ and is indeed capacity achieving, \cite {our_journal2025} with $\mathscr{C}(\mathbb{P}_N) = \frac{2}{N}$. Specifically, let the PIR system corresponding to $\mathbb{P}_N$ consist of the set of files $\{W_1, \ldots, W_{N-1}\}$, where $S_1$ stores $W_1$, $S_N$ stores $W_{N-1}$, and $S_i$, $i \in [2:N-1]$, stores $\{W_{i-1}, W_i\}$. Suppose each $W_i$ consists of $2$ bits, which the user randomly and independently permutes into $w_i=(w_i(1),w_i(2))$. To retrieve $W_{\theta}$, the user downloads, $w_1(1)$ from $S_1$, $w_{i-1}(1)+w_{i}(1)$ from $S_i, i\in [2:\theta]$, $w_{i-1}(2)+w_{i}(2)$ from $S_i, i\in [\theta+1:N-1]$, and $w_{N-1}(2)$ from $S_N$. 
    
    The above scheme requires $N$ downloaded bits in total, one from each server. The user retrieves $w_{\theta}(1)$ from $S_{\theta}$ and $w_{\theta}(2)$ from $S_{\theta+1}$. Clearly, the scheme satisfies the SRP and results in rate $\frac{2}{N}$. Thus, by Theorem \ref{thm:achievable_lbnd}, this leads to
    \begin{align}
        \mathscr{C}\left(\mathbb{P}_N^{(r)}\right)\geq \frac{2}{N} \cdot  \left(\frac{1}{2-\frac{1}{2^{r-1}}} \right). \label{eq_lower boud_path}
    \end{align}
\end{remark}

\begin{figure}[t]
    \centering
    \includegraphics[width=0.4\textwidth]{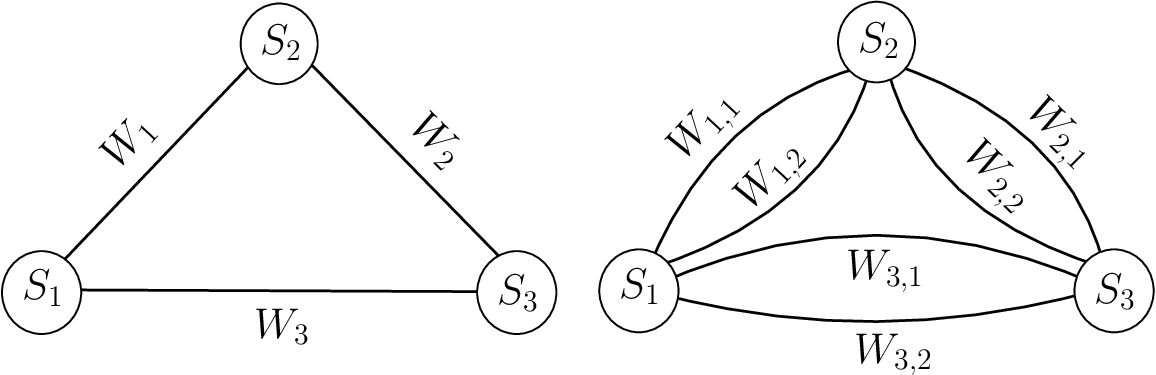}
    \caption{PIR systems for cycle $\mathbb{C}_3$ (left) and $2$-multicycle $\mathbb{C}_3^{(2)}$ (right).}
    \label{fig:2-multicycle_PIR}
\end{figure}

\begin{remark}
    Consider the PIR system arising from multicycles $\mathbb{C}_N^{(r)}$, as shown in Fig.~\ref{fig:2-multicycle_PIR}. The capacity-achieving PIR scheme for $\mathbb{C}_N$ proposed in \cite{BU19} satisfies the SRP, and the capacity is given by $\mathscr{C}(\mathbb{C}_N) = \frac{2}{N+1}$. Thus, by  Theorem \ref{thm:achievable_lbnd}, we also have
    \begin{align}
        \mathscr{C}\left(\mathbb{C}_N^{(r)}\right)\geq \frac{2}{N+1}\cdot \left(\frac{1}{2-\frac{1}{2^{r-1}}} \right).
    \end{align}
\end{remark}

\section{An Upper Bound on the PIR Capacity}
For a graph $G = (V, E)$, let $\Delta(G)$ denote the \emph{maximum degree} of $G$, and let $I(G)$ denote its \emph{incidence matrix}. A \emph{matching} $ M \subseteq E $ of $ G $ is a subset of edges such that no two edges share a common vertex. The \emph{matching number} $ \nu(G) $ of $ G $ is the maximum size of a matching in $ G $.

\begin{theorem}\label{Thm: Capacity upper bound for multigraphs}
    Let $ G = (\cS, \cW_0) $ be a graph with maximum degree $ \Delta(G) $. Then, the PIR capacity of $G^{(r)}$ satisfies:
    \begin{align}
        \mathscr{C}\left(G^{(r)}\right) \leq \min \left( \frac{\Delta(G)}{|E(G)|}, \frac{1}{\nu(G)} \right)\cdot\frac{1}{2-2^{1-r}}.
    \end{align}
\end{theorem}

\begin{remark}\label{Connection between UP bounds of graphs and multigraphs}
    When $ r = 1 $, the result of Theorem \ref{Thm: Capacity upper bound for multigraphs} coincides with that of Theorem 1 in \cite{SGT23}. The capacity upper bound of Theorem \ref{Thm: Capacity upper bound for multigraphs} is tight for certain graphs. For example, set $ G = \mathbb{P}_N $ with $ 2\mid N $, by Theorem \ref{Thm: Capacity upper bound for multigraphs}, 
    \begin{align*}
        \mathscr{C}\left(\mathbb{P}_N^{(r)}\right) &\leq \min \left( \frac{2}{N-1}, \frac{1}{N/2} \right) \cdot \frac{1}{2 - 2^{1-r}} \\
        &= \frac{2}{N} \cdot \frac{1}{2 - 2^{1-r}}, 
    \end{align*}
    which matches the lower bound in \eqref{eq_lower boud_path}.
\end{remark}

For the proof of Theorem \ref{Thm: Capacity upper bound for multigraphs}, we need the following multigraph version of Lemma 3 in \cite{SGT23}. 

\begin{lemma}\label{multigraph_version_Lem3}
    Let $i,j\in [N]$ be two distinct servers that share a set of $r$ files $\cW_{i,j}\triangleq\{W_1,W_2,\ldots,W_r\}\subseteq \cW$, then
    \begin{align}
    H(A_i)\!+\!H(A_j) &\geq  H(A_i|\cW\setminus\cW_{i,j},\mathcal{Q})\!+\!H(A_j|\cW\setminus\cW_{i,j},\mathcal{Q}) \nonumber \\
    & \geq \left(1+\frac{1}{2}+\cdots+\frac{1}{2^{r-1}}\right)L.
    \end{align}
\end{lemma}
The proof of Lemma~\ref{multigraph_version_Lem3} requires the following extension of \cite[Lemma~6]{SJ17}, whose proof follows similarly.

\begin{lemma}\label{Lem6 in Sun&Jafar's paper under graph PIR model}
    Let $i,j\in [N]$ be two distinct servers that share a set of $r$ files $\cW_{i,j}\triangleq\{W_1,W_2,\ldots,W_r\}\subseteq \cW$, then for any $s\in \{2,\ldots,r\}$, it holds that
    \begin{align*}
        & I(\cW_{i,j}\setminus W_{[s-1]}; A_i,A_j|\cW\setminus\cW_{i,j}\cup W_{[s-1]}, \cQ,\theta=s-1) 
        \\
        &\geq   \frac{L}{2}+\frac{1}{2}\cdot I(\cW_{i,j}\setminus W_{[s]}; A_i,A_j|\cW\setminus\cW_{i,j}\cup W_{[s]}, \cQ,\theta=s),
    \end{align*}
    where $W_{[s]}\triangleq\{W_1,W_2,\ldots,W_s\}$.
\end{lemma}

\begin{Proof}[\emph{Proof of Lemma~\ref{multigraph_version_Lem3}}]
Let $\cW^c$ denote $\cW\setminus \cW_{i,j}$ and $\Delta$ be the difference of the following entropies, 
    \begin{align}
        \Delta \triangleq & H(\cW_{i,j}|\cW^c, \mathcal{Q},\theta=1)
        \nonumber \\
        & - H(\cW_{i,j}|\cW^c, \mathcal{Q},\theta=1,A_{[N]})
        \label{eq_Delta_def}.
    \end{align}
    Then,
    \begin{align}
        \Delta = &H(\cW_{i,j}|\cW^c, \mathcal{Q},\theta=1) \nonumber \\
        &-H(\cW_{i,j}|\cW^c, \mathcal{Q},\theta=1,A_{i},A_{j})
         \label{eq1_Delta} \\
        =&I(\cW_{i,j}; A_{i},A_{j}|\cW^c, \mathcal{Q},\theta=1) \\
        =&H(A_{i},A_{j}|\cW^c, \mathcal{Q},\theta=1) \nonumber \\
        &- H(A_{i},A_{j}|\cW, \mathcal{Q},\theta=1) \\
        =&H(A_{i},A_{j}|\cW^c, \mathcal{Q},\theta=1) \label{eq2_Delta} \\
        \leq & H(A_{i}|\cW^c, \mathcal{Q},\theta=1) \nonumber\\
        &+H(A_{j}|\cW^c, \mathcal{Q},\theta=1) \\
        = &H(A_{i}|\cW^c, \mathcal{Q})+H(A_{j}|\cW^c, \mathcal{Q}), \label{eq3_Delta}
    \end{align}
    where \eqref{eq1_Delta} and \eqref{eq2_Delta} follow by \eqref{eq3_problem_setting}, and \eqref{eq3_Delta} follows by \eqref{eq5_problem_setting}; see also \cite[Proposition~2]{SGT23}. On the other hand, we have
    \begin{align}
        \Delta = &rL-H(\cW_{i,j}|\cW^{c}, \mathcal{Q},\theta=1,A_{i},A_{j})
         \label{eq4_Delta} \\
        =&rL-H(W_1|\cW^{c}, \mathcal{Q},\theta=1,A_{i},A_{j}) \nonumber\\
        & -H(\cW_{i,j}\setminus\{W_1\}|\cW^{c}\cup W_1, \mathcal{Q},\theta=1,A_{i},A_{j}) \\
        =&rL-H(W_1|\cW^{c}, \mathcal{Q},\theta=1,A_{[N]}) \nonumber \\
        & -H(\cW_{i,j}\setminus\{W_1\}|\cW^{c}\cup W_1, \mathcal{Q},\theta=1,A_{i},A_{j})
         \label{eq5_Delta} \\
        =&rL \nonumber\\
        & -H(\cW_{i,j}\setminus\{W_1\}|\cW^{c}\cup W_1, \mathcal{Q},\theta=1,A_{i},A_{j})
         \label{eq6_Delta} \\ 
        =&rL-H(\cW_{i,j}\setminus\{W_1\}|\cW^{c}\cup W_1, \mathcal{Q},\theta=1) \nonumber \\
        & +I(\cW_{i,j}\setminus\{W_1\}; A_i,A_j|\cW^{c}\cup W_1, \mathcal{Q},\theta=1)\\ 
        =&L+I(\cW_{i,j}\setminus\{W_1\}; A_i,A_j|\cW^{c}\cup W_1, \mathcal{Q},\theta=1),
         \label{eq7_Delta}
    \end{align}
    where \eqref{eq4_Delta} and \eqref{eq7_Delta} follow by \eqref{eq1_problem_setting} and \eqref{eq2_problem_setting}, \eqref{eq5_Delta} follows since answers in $A_{[N]}\setminus \{A_i,A_j\}$ are functions of $\mathcal{Q}$ and $\cW^{c}$, and \eqref{eq6_Delta} follows by the reliability \eqref{eq4_problem_setting}.

    Next, starting from $s=2$ and applying Lemma \ref{Lem6 in Sun&Jafar's paper under graph PIR model} repeatedly for $s=3$ to $r$, we obtain that
    \begin{align}
         I(&\cW_{i,j}\setminus\{W_1\}; A_i,A_j|\cW^{c}\cup W_1, \mathcal{Q},\theta=1) 
         \nonumber \\
        &\geq  \frac{L}{2}+\frac{1}{2}\cdot I(\cW_{i,j}\setminus \cW_{[2]}; A_i,A_j|\cW^{c}\cup \cW_{[2]}, \mathcal{Q},\theta=2) 
         \nonumber \\
        &\geq  \frac{L}{2}+\frac{L}{4}+\frac{1}{4}\cdot I(\cW_{i,j}\setminus \cW_{[3]}; A_i,A_j|\cW^{c}\cup \cW_{[3]}, \mathcal{Q},\theta=3)
         \nonumber \\
        &\geq  \cdots 
        \geq \left(\frac{1}{2}+\frac{1}{4}+\cdots+\frac{1}{2^{r-1}}\right)L.
         \label{eq13_Delta}
    \end{align}
    Then, the result follows by combining \eqref{eq7_Delta} and \eqref{eq13_Delta}.
\end{Proof}

With Lemma~\ref{multigraph_version_Lem3}, we can proceed to prove Theorem~\ref{Thm: Capacity upper bound for multigraphs}.

\begin{Proof}[\emph{Proof of Theorem~\ref{Thm: Capacity upper bound for multigraphs}}]
Consider the rate of some PIR scheme $T$ for $ G^{(r)} $,
\begin{align}
    R = \frac{L}{\sum_{i \in [N]} H(A_i)} 
    = \frac{1}{\sum_{i \in [N]} \left(H(A_i)/L\right)} 
    = \frac{1}{\mathbf{1}_N \cdot \mu^T},
\end{align}
where $ \mu_i \triangleq \frac{H(A_i)}{L} $ for each $ i \in [N] $, $ \mu \triangleq (\mu_1, \ldots, \mu_N) $, and $ \mathbf{1}_N $ is the all-ones row vector of length $ N $. According to Lemma~\ref{multigraph_version_Lem3}, if servers $ i $ and $ j $ are adjacent, then 
\begin{align}
    \mu_i + \mu_j\geq 1+\frac{1}{2}+\cdots+\frac{1}{2^{r-1}}=2-\frac{1}{2^{r-1}}.
\end{align}
Hence, we can get an upper bound on $R$ via the reciprocal of the optimal value of the following linear program,
\begin{align}
     \min & \quad \mathbf{1}_N \cdot \mu^T \notag \\
    \text{s.t.} & \quad I(G)^T \cdot \mu^T \geq \left( 2-\frac{1}{2^{r-1}} \right) \cdot \mathbf{1}_{K'} \notag \\
    & \quad \mu_i \geq 0, \ i\in [N],
\end{align}
where $ I(G) $ is the incidence matrix of graph $ G $ and $K'=|\cW_0|$. Its dual problem is
\begin{align}
    \max & \quad \left( 2-\frac{1}{2^{r-1}} \right) \cdot\mathbf{1}_{K'} \cdot \eta^T \notag \\
    \text{s.t.} &  \quad I(G) \cdot \eta^T \leq \mathbf{1}_N \notag\\
    & \quad \eta_i  \geq 0, \ i\in [K']. \label{eq_dual_problem}
\end{align}
By the primal-dual theory, any feasible solution to the dual problem provides a lower bound for the primal problem. Below, we provide two feasible solutions to the dual problem:
\begin{itemize}
    \item The vector $ \bbv_1 = \frac{1}{\Delta} \cdot \mathbf{1}_{K'} $. Thus, by $\mathbf{1}_{K'} \cdot \bbv_1^T = \frac{K'}{\Delta}$,
    the rate $R$ is at most $ \Delta/K' \cdot \left(2-2^{1-r} \right)^{-1}$.
    \item The indicator vector $ \bbv_2 \in \{0, 1\}^{K'} $ of a maximum matching $ M \subseteq \cW_0 $ of $ G $, i.e., $ |M| = \nu(G) $. Thus, the rate $R$ is at most $ 1/\nu(G)\cdot \left(2-2^{1-r} \right)^{-1} $.
\end{itemize}
This completes the proof.    
\end{Proof}

\section*{Acknowledgment}
This work was supported in part by the European Research Council (ERC) under Grant 852953. 

\newpage

\bibliographystyle{IEEEtran}
\bibliography{biblio_new}

\end{document}